\documentclass[twocolumn,aps,prl]{revtex4}
\usepackage{graphicx}
\begin{document}
\title{Magnetic Penetration Depth Measurements of 
       Pr$_{2-x}$Ce$_x$CuO$_{4-\delta}$ Films on Buffered Substrates:
       Evidence for a Nodeless Gap}
\author{Mun-Seog Kim}
\author{John A. Skinta}
\author{Thomas R. Lemberger}
\affiliation{Department of Physics, 
             Ohio State University, 
             Columbus, 
             OH 43210-1106}
\author{A. Tsukada}
\author{M. Naito}
\affiliation{NTT Basic Research Laboratories, 
	 3-1 Morinosato Wakamiya, 
	 Atsugi-shi, Kanagawa 243, Japan}

\begin{abstract}
We report measurements of the inverse squared magnetic penetration depth, 
$\lambda^{-2}(T)$, in Pr$_{2-x}$Ce$_{x}$CuO$_{4-\delta}$ 
($0.115 \leq x \leq 0.152$) superconducting films grown on SrTiO$_3$ 
(001) substrates coated with a buffer layer of insulating Pr$_{2}$CuO$_{4}$. 
$\lambda^{-2}(0)$, $T_c$ and normal-state resistivities of these films 
indicate that they are clean and homogeneous. 
Over a wide range of Ce doping, $0.124\leq x \leq 0.144$,
$\lambda^{-2}(T)$ at low $T$ is flat: it changes by less than 0.15\% 
over a factor of 3 change in $T$, indicating a gap in the superconducting 
density of states. Fits to the first 5\% decrease in $\lambda^{-2}(T)$ 
produce values of the minimum superconducting gap in the range 
of $0.29\leq\Delta_{\rm min}/k_BT_c\leq1.01$.
\end{abstract}
\pacs{PACS number(s):}
\maketitle

It is still a puzzle whether pairing symmetry in $n$-type cuprates is $d$ wave 
or not\cite{tsuei2,armitage1,kokales1,prozorov1,alff2,kashiwaya1, ctchen1,alff1,skinta1}. 
Recently, novel concepts on pairing symmetry of $n$- and $p$-type cuprates have come forward: 
a possible transition in pairing symmetry\cite{skinta2,Biswas1} 
and/or a mixed symmetry order parameter\cite{muller1,kohen1,daghero1}. 
Our previous work\cite{skinta2} involved La$_{2-x}$Ce$_x$CuO$_{4-\delta}$ (LCCO) 
and Pr$_{2-x}$Ce$_x$CuO$_{4-\delta}$ (PCCO) films grown directly on SrTiO$_3$ 
substrates. We found that at low Ce doping levels, $\lambda^{-2}(T)$ at low $T$ was 
quadratic in $T$, but at higher dopings, $\lambda^{-2}(T)$ showed activated 
behavior. These results suggested a $d$- to $s$-wave pairing transition near optimal 
doping, as was also suggested by tunneling experiments\cite{Biswas1} on PCCO films. 
We have subsequently improved film quality by eliminating the interface between the film and 
substrate, by growing PCCO films onto Pr$_2$CuO$_4$ 
(PCO)/SrTiO$_3$ instead of directly onto SrTiO$_3$. The insulating PCO layer is 
thought to lessen lattice mismatch between PCCO film and SrTiO$_3$ substrate,  
so that these films should be more homogeneous through their thickness. 
In fact, their normal state resistivities are somewhat lower than those of 
unbuffered PCCO films for the same doping, $x$. $T_c$'s at optimal 
doping in the two film families are the same, $T_c\simeq 24$ K.

Films were prepared by molecular-beam epitaxy (MBE) on 10 mm
$\times$ 10 mm $\times$ 0.35 mm SrTiO$_{3}$ substrates as detailed 
elsewhere\cite{naito01}.
The same growth procedures and
parameters were used for all films. For all films,
PCCO and PCO layers are 750 \AA~and 250 \AA~thick,
respectively. Ce concentrations, $x$, are measured to better than 
$\pm 0.005$ by inductively coupled plasma spectroscopy. X-ray rocking 
curves show full-width at half maximum of (006) reflection for all
films to be less than 0.4$^\circ$, which implies that the films are 
highly $c$-axis oriented.

The penetration depth, $\lambda(T)$, was measured down to $T\simeq 0.5$ K 
using a mutual inductance apparatus, described in detail 
elsewhere\cite{turneaure_1,turneaure_2}, in a He$^3$ refrigerator.
The system temperature was measured with a Cernox resistor  
(LakeShore Inc.) and its reliability, below 1 K, was confirmed 
by measuring the superconducting transition temperature of a 
Zn plate, $T_c=0.875$ K.

Each film was centered between drive and pick-up coils with diameters 
of $\sim 1$ mm. A small current at 50 kHz in the drive coil induced diamagnetic
screening currents in the film, i.e., parallel to the CuO$_2$ planes.
The time derivative of the net magnetic field from drive coil and 
induced current in the film was measured as a voltage across 
the pick-up coil. The real and imaginary parts of the mutual 
inductance are proportional to the quadrature and in-phase 
components of ac voltage, respectively.
Because the coils were much smaller than the film, the applied field was
concentrated near the center of the films, and demagnetizing effects at the film 
perimeter were not relevant. Because films were thinner than $\lambda$, the 
current density induced in the films was essentially uniform through the 
film thickness. Nonlinear effects occur only very close to $T_c$ 
where $\lambda^{-2}$ is less than 1\% of its value at $T=0$. 
All data presented here represent linear response. 

The procedure to extract $\lambda^{-2}(T)$ from the measured mutual inductance 
is the following.
First, a constant background mutual inductance due to stray couplings between 
drive and pickup circuits is subtracted from raw data. This background is 
the measured mutual inductance at $4.2$ K with the sample replaced by a 
100 micron-thick superconducting Pb foil with identical shape and area. 
No magnetic field passes through the Pb foil. A glass shim ensures that 
the spacing between coils is the same as with the real sample. 
The adjusted data are normalized to the mutual inductance measured 
above $T_c$, at $T \sim 30$ K, where the film is utterly transparent to the ac field. 
Normalization reduces uncertainties associated with amplifier gains 
and nonideal aspects of the coil windings.
A numerical model of the drive and pick-up coils enables us to convert 
the subtracted and normalized mutual inductance to 
$sheet$ conductivity: $\sigma d=\sigma_1 d-i\sigma_2 d$, 
where $d$ is film thickness.
Finally, $\lambda^{-2}$ is determined from $\sigma_2 d$ via: 
$\sigma_2=1/\mu_0\omega\lambda^2$, where $\mu_0$ is the 
magnetic permeability of vacuum and $\omega$ is the angular 
frequency of the drive current. The absolute accuracy 
of $\lambda^{-2}$ is limited by $\pm 10$\% uncertainty in $d$. 
The $T$ dependence of $\lambda^{-2}$ is unaffected by this uncertainty. 

\begin{figure}[tb]
\centerline{\includegraphics[height=2.4in,angle=0]{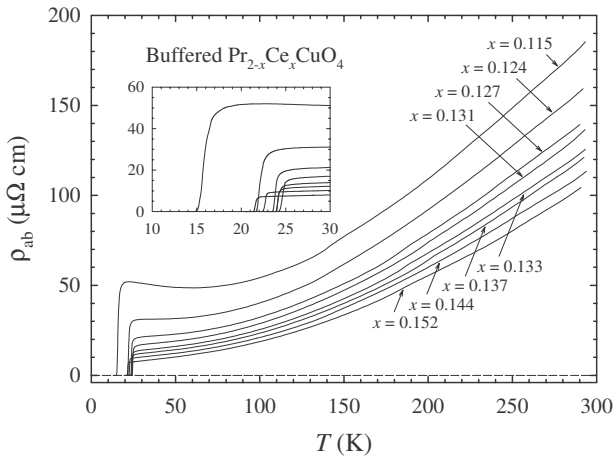}}
\caption{$ab$-plane resistivities, $\rho_{ab}(T)$, of 
	 buffered Pr$_{2-x}$Ce$_x$CuO$_{4-\delta}$ films.
	 For resistivities at $T=25$ K, see Table I. 
	 Inset: $\rho_{ab}(T)$ around $T_c$.}
\label{fig1}
\end{figure}

%results
Except for differences in the flatness of $\lambda^{-2}$ at low $T$, which 
is the focal point of this paper, buffered films are very much like 
unbuffered films reported earlier\cite{skinta2}. Fig.\ \ref{fig1} 
shows in-plane resistivity, $\rho_{ab}(T)$, for buffered PCCO films. 
$\rho_{ab}$ in the normal state decreases smoothly and monotonically 
with Ce doping, $x$, even for small changes in $x$, implying that 
the main difference among films is Ce content. If there were random 
variations in degree of epitaxy, structural defects, etc., then 
resistivity would not be such a smooth function of $x$. These 
resistivities are slightly smaller than for PCCO films without 
buffer layers\cite{skinta1,skinta2}, and significantly lower than for NCCO and PCCO 
crystals\cite{kokales1,kokales2}. The inset of Fig.\ \ref{fig1} 
shows that resistive transitions are reasonably sharp, and 
that $T_c$ is a weak function of Ce concentration, although resistivity is not. 
Table\ \ref{table1} summarizes properties of the films.

\begin{figure}[tb]
\centerline{\includegraphics[height=2.4in,angle=0]{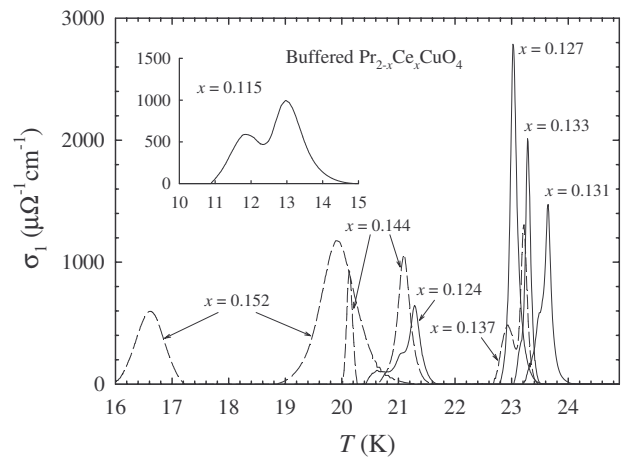}}
\caption{$\sigma_1(T)$ at 50 kHz in buffered Pr$_{2-x}$Ce$_x$CuO$_{4-\delta}$ films. 
	 Inset: $\sigma_1(T)$ for the film with $x=0.115$.}
\label{fig2}
\end{figure}

Fluctuations cause $\sigma_1(T)$ to peak at the superconducting transition. 
Hence, $\sigma_1(T)$ is a much more stringent test of film quality 
than resistivity. For example, if $T_c$ varies through the film thickness, 
resistivity reveals only the highest $T_c$. Because our probing magnetic 
field passes through the film, $\sigma_1(T)$ has a peak at the $T_c$ of 
every layer. Transitions associated with small bad spots in the film, as opposed 
to an entire film layer, are distinguished by their having essentially 
no effect on the superfluid response, $\sigma_2$. When a layer goes 
superconducting there is a distinct change in the slope of $\lambda^{-2}(T)$.

\begin{figure}[tb]
\centerline{\includegraphics[height=2.4in,angle=0]{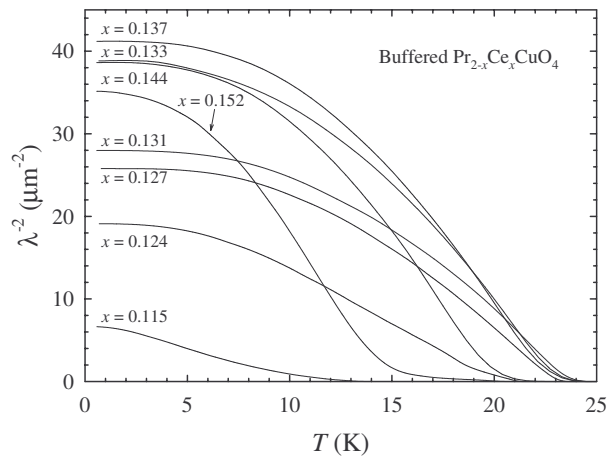}}
\caption{$\lambda^{-2}(T)$ for buffered Pr$_{2-x}$Ce$_x$CuO$_{4-\delta}$ films. 
         Film-to-film uncertainty in $\lambda^{-2}(0)$ is $\sim \pm 10$\%. }
\label{fig3}
\end{figure}

$\sigma_1(T)$'s of buffered PCCO films (Fig.\ \ref{fig2}) show that several 
of them have a double transition, reflected as shoulder ($x=$ 0.115, 0.124, and 0.137) 
or satellite ($x=$ 0.144 and 0,152) structure of peaks. We define two transition 
temperatures, $T_{c1}$ and $T_{c2}$, from peaks 
in $\sigma_{1}(T)$, where $T_{c1} > T_{c2}$. The resistive $T_c$ is always 
at the onset of the $T_{c1}$-peak. 
For the films most important to the conclusions of this paper, 
$0.124\leq x \leq 0.144$, the width of the $T_{c1}$ peak, $\Delta T_{c1}$, is $\leq 1$ K, 
indicating excellent film homogeneity. The peak at $T_{c2}$ most likely 
involves a bad spot in the film, since there is no corresponding feature 
in the slope of $\lambda^{-2}(T)$, (see Fig.\ \ref{fig3}). Accordingly, 
the lower transition is neglected in our analysis. Films with highest and 
lowest Ce concentrations ($x=$ 0.115 and 0.152) 
have broader transition widths ($\Delta T_c = 2.4 \sim 3.9$ K) than 
other films, perhaps because $T_c$ is more sensitive to $x$.

Figure\ \ref{fig3} shows $\lambda^{-2}(T)$ for all films. $\lambda^{-2}(0)$ vs. 
$x$ increases rapidly for $x\leq 0.133$, and it is constant or decreases 
slowly for $x > 0.133$. Values of $\lambda^{-2}(0)$ are slightly higher 
than for unbuffered films. The surprising upward curvature that develops in 
$\lambda^{-2}(T)$ near $T_c$ at high Ce concentrations was also observed in 
unbuffered LCCO and PCCO films\cite{skinta1,skinta2}. 

In our previous work\cite{skinta2} on unbuffered PCCO films, films 
with low Ce concentrations showed quadratic ($T^2$) behavior 
in $\lambda^{-2}(T)$ at low $T$. Films with high Ce concentrations 
showed gap-like behavior:  
\begin{equation}
\lambda^{-2}(T) \simeq \lambda^{-2}(0) [ 1 - C_{\infty} \exp(-D/t) ],
\label{expfor}
\end{equation}
where $C_{\infty}$ and $D$ are adjustable parameters, and $t=T/T_c$. 
In the clean limit, $D$ is approximately the minimum gap on the Fermi surface, 
normalized to $k_B T_c$, and $C_{\infty}$ is roughly twice the average 
superconducting density of states (DOS) over energies within $k_B T$ of the 
gap edge. For isotropic BCS superconductors, the best-fit value of 
$C_\infty/2$ is about 2.2.
The change in low-$T$ behavior of $\lambda^{-2}(T)$ near optimal 
doping suggested a transition in pairing symmetry. 

\begin{figure}[tb]
\centerline{\includegraphics[height=4in,angle=0]{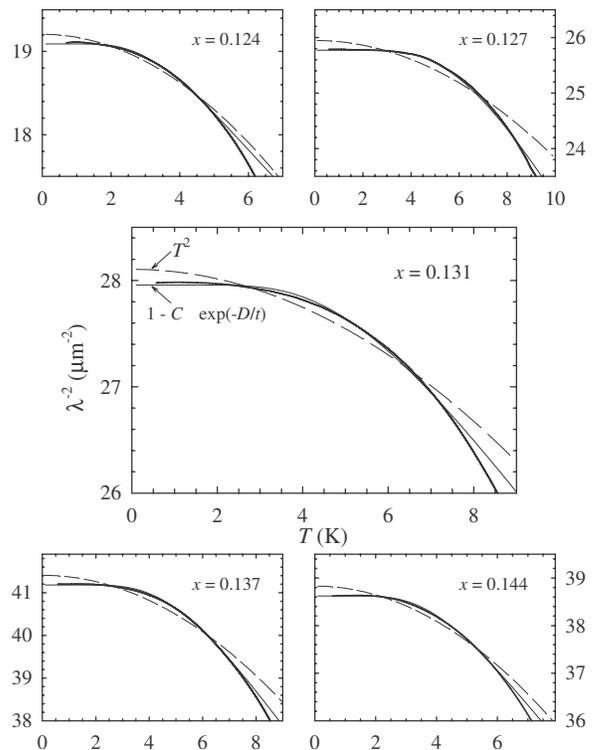}}
\caption{Expanded view of $\lambda^{-2}(T)$ at low $T$ for 
	 buffered Pr$_{2-x}$Ce$_x$CuO$_{4-\delta}$ films. 
	 Thin solid and dashed lines denote best fits of 
	 $\lambda^{-2}(0)[1-C_{\infty}\exp(-D/t)]$ and $\lambda^{-2}(0)[1-(T/T_0)^2]$ 
	 to the first 5\% drop in $\lambda^{-2}(T)$, respectively.}
\label{fig4}
\end{figure}

We now turn to the low-$T$ behavior of $\lambda^{-2}(T)$ for 
buffered PCCO films, shown on a greatly expanded scale 
in Fig.\ \ref{fig4}. The most important thing to notice is that $\lambda^{-2}(T)$ 
is flat to better than 0.15\% over a factor of 3 or more change 
in $T$. Residual variations in $\lambda^{-2}(T)$ at the 0.1\% 
level are due, at least in part, to slow drift in the gain of 
the lock-in amplifiers used to measure current and voltage. 
These data are incompatible with simple $d$-wave models with nodes 
in the gap. Thus, except for the most underdoped and overdoped films 
($x=$ 0.115 and 0.152), $\lambda^{-2}(T)$ shows gapped behavior. 
Recent angle-resolved photoemission spectroscopy measurements\cite{armitage1} 
indicate well-defined quasiparticle states on the Fermi surface where 
the $d_{x^2-y^2}$ node would be, so the gapped behavior that we observed 
could not be ascribed to a Fermi surface effect.

\begin{figure}[tb]
\centerline{\includegraphics[height=2.4in,angle=0]{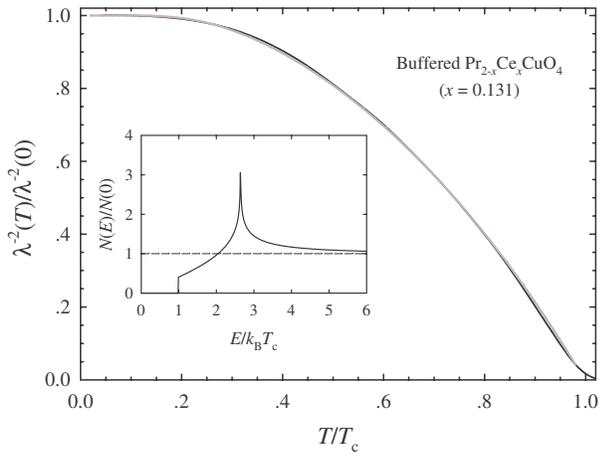}}
\caption{$\lambda^{-2}(T)$ for Pr$_{1.869}$Ce$_{0.131}$CuO$_{4-\delta}$ film. 
Gray line shows an excellent fit obtained with density of states 
shown in the inset. Inset: Quasiparticle density of states in $s+id_{x^2-y^2}$ 
gap symmetry.} 
\label{fig5}
\end{figure}

To estimate the gap, we fit Eq. (\ref{expfor}) to the first 
$\sim 5$\% drop in $\lambda^{-2}(T)$, (thin solid lines in Fig.\ \ref{fig4}). 
It comes as no surprise that quadratic fits over the same temperature 
range are unacceptable (dashed lines). For films with $x=$ 0.115 and 0.152, 
data at $T<0.5$ K are needed to distinguish between $T^2$ and $e^{-D/t}$. 
Values of the minimum gap, $\Delta_{\rm min}=Dk_BT_c$ and average DOS, 
$C_{\infty}/2$, extracted from the above exponential fits 
are presented in Table\ \ref{table1}. $D$ values are significantly lower than 
the BCS weak-coupling-limit value, $1.76$, for $s$-wave superconductors 
(2.14 for $d$-wave superconductors). 
$D$ is largest, $D \sim 1$, for $x$ near 0.13. A similar value, 
$D \simeq 0.85$, was found for unbuffered PCCO films \cite{skinta1} with the 
same Ce concentration. Values of $C_\infty/2$ ($\ll 1$) are also much smaller 
than for weak-coupling isotropic $s$ wave. This implies existence of a peak 
in the DOS for a certain $E$ ($>\Delta_{\rm min}$), because the states 
should be conserved. 

The next question is: where is the peak in the DOS, 
i.e., how big is the maximum gap, $\Delta_{\rm max}$, on the Fermi surface? To answer 
this question, we employ a model anisotropic gap function and the 
clean-limit result that $1-\lambda^{-2}(T)/ \lambda^{-2}(0)$ is an 
integral of quasiparticle DOS times the derivative 
of the Fermi function with respect to energy\cite{tinkham1}. 
Fig.\ \ref{fig5} shows a good fit to $\lambda^{-2}(T)$ for film with $x=0.131$  
using the DOS in the inset. 
In this fit, the minimum gap was fixed at the 
value found by fitting the low-$T$ data, i.e., $\Delta_{\rm min}/k_BT_c=0.99$.
Then, as one can see in inset of Fig.\ \ref{fig5}, the average DOS 
within $\sim k_B T$ of the minimum gap edge agrees well with $C_\infty/2=0.5$ 
from Table\ \ref{table1}. For film with $x=0.131$, $\Delta_{\rm max}$ is 
about $2.6k_BT_c ~(\pm 15\%)$. 

We emphasize that 
we cannot say anything about the shape of the peak in the DOS, 
only its location. An equally acceptable fit, with a 
similar peak energy, is obtained even when the sharp narrow peak 
in the inset of Fig.\ \ref{fig5} is replaced by a 
rectangular peak\cite{skinta3}. 

In summary, we measured the inverse squared magnetic 
penetration depth, $\lambda^{-2}(T)$, of several 
Pr$_{2-x}$Ce$_x$CuO$_{4-\delta}$ films on buffered 
Pr$_2$CuO$_4$/SrTiO$_3$ substrates down to $T/T_c < 0.03$. 
Overall, the resistivities and penetration depths were similar 
to films grown directly on SrTiO$_3$. However, for PCCO films 
on buffered substrates, $\lambda^{-2}(T)$ at low $T$ exhibits 
gapped behavior over a wide range of Ce doping, including 
underdoping. This implies a superconducting gap without 
nodes on the Fermi surface. Values of the minimum 
superconducting gap for the films 
are in range of $0.3 \leq \Delta_{\rm min}/k_BT_c \leq 1.0$. 
We cannot distinguish among models with various gap symmetries, 
e.g., anisotropic $s$,  $s+id$, or $d+id$.

The research at OSU was supported by NSF Grant No. DMR-0203739.

\begin{table}
\caption{Properties of eight MBE-grown Pr$_{2-x}$Ce$_x$CuO$_{4-\delta}$ films on 
         Pr$_2$CuO$_4$/SrTiO$_3$. $T_{c}$ (or $T_{c1}$) and $T_{c2}$ are 
         locations of main and secondary peaks in $\sigma_{1}(T)$, respectively.
         $\Delta T_c$ is full width of the (main) peak in $\sigma_{1}(T)$.
         $\rho_{ab}$(25 K) is the $ab$-plane
         resistivity at $T=25$ K. $\lambda^{-2}(0)$, $C_{\infty}/2$, 
         and $D=\Delta_{\rm min}/k_BT_c$ 
         are fit parameters, in Eq.\ (\ref{expfor}).}
\begin{ruledtabular}
\begin{tabular}{cccccccc}
$x$ & $T_c$ $(T_{c1})$& $T_{c2}$&$\Delta T_c$&$\rho_{ab}$(25 K)  &  $\lambda^{-2}(0)$ & $C_{\infty}/2$ & $D$ \\
 &  (K) & (K)& (K) &($\mu\Omega$cm) & ($\mu\rm{m}^{-2}$) & &  \\
\hline \hline
0.115 &  13.0 &11.8& 3.9&51.0 &  6.6 &(0.21)&(0.29)\\
0.124 &  21.3 &20.7& 1.3&30.1 & 19.1 & 0.28 & 0.56 \\
0.127 &  23.1 &    & 0.8&19.4 & 25.8 & 0.60 & 1.01 \\

0.131 &  23.6 &    & 0.8&15.3 & 27.9 & 0.50 & 0.99 \\
0.133 &  23.3 &    & 0.5&12.8 & 38.9 & 0.42 & 0.73 \\
0.137 &  23.2 &22.9& 0.7&10.8 & 41.2 & 0.38 & 0.83 \\
0.144 &  21.2 &20.2& 0.9&9.5  & 38.6 & 0.30 & 0.72 \\
0.152 &  19.8 &16.6& 2.4&7.7  & 35.1 &(0.17)&(0.37)\\
\end{tabular}
\end{ruledtabular}
\label{table1}
\end{table}

\end{document}